\documentclass[11pt]{amsart}

\usepackage{amsfonts}
\usepackage{amsmath}
\usepackage{amssymb}

\theoremstyle{plain}
\newtheorem{theorem}{Theorem}[section]

\theoremstyle{definition}

\theoremstyle{remark}

\numberwithin{equation}{section}

\begin{document}
\thispagestyle{empty}
\begin{center}
\null\vspace{-1cm}
{\footnotesize Available at: 
{\tt http://publications.ictp.it}}\hfill IC/2007/109\\
\vspace{0.5cm}
United Nations Educational, Scientific and Cultural Organization\\
and\\
International Atomic Energy Agency\\
\medskip
THE ABDUS SALAM INTERNATIONAL CENTRE FOR THEORETICAL PHYSICS\\
\vspace{2cm}
{\bf ANALYTICAL PROPERTIES OF AN OSTROVSKY-WHITHAM TYPE DYNAMICAL SYSTEM FOR A RELAXING MEDIUM WITH SPATIAL MEMORY AND ITS INTEGRABLE REGULARIZATION}\\
\vspace{1.7cm}
Nikolai N. Bogoliubov (Jr.)\footnote{nikolai\_bogolubov@hotmail.com}\\
{\it V.A. Steklov Mathematical Institute of RAN, Moscow, Russian Federation\\
and\\ 
The Abdus Salam International Centre for Theoretical Physics, Trieste,
Italy,}\\[1.5em]
Anatoliy K. Prykarpatsky\footnote{prykanat@cybergal.com}\\
{\it The Department of Nonlinear Mathematical Analysis at the IAPMM of
National Academy of Sciences, The AGH University of Science and
Technology, Krakow 30059, Poland\\
and\\
The Abdus Salam International Centre for Theoretical Physics, Trieste,
Italy,}\\[1.5em]
Ilona Gucwa\footnote{ilonagucwa@interia.pl} and Jolanta Golenia\footnote{goljols@tlen.pl}\\
{\it Department of Applied Mathematics, The AGH University of Science and Technology,\\ Krakow 30059, Poland.}
\end{center}
\vspace{0.5cm}
\centerline{\bf Abstract}
\baselineskip=18pt
\bigskip

Short-wave perturbations in a relaxing medium, governed by a
special reduction of the Ostrovsky evolution equation, and later
derived by Whitham, are studied using the gradient-holonomic
integrability algorithm.The bi-Hamiltonicity and complete
integrability of the corresponding dynamical system is stated and
an infinite hierarchy of  commuting to each other conservation
laws of dispersive type  are found. The two- and four-dimensional
invariant reductions are studied in detail. The well defined
regularization of the model is constructed and its Lax type
integrability is discussed.
\vfill
\begin{center}
MIRAMARE -- TRIESTE\\
December 2007\\
\end{center}
\vfill


\newpage

\section{Introduction}

Many important problems of propagating waves in nonlinear media with
distributed parameters can be described by means of evolution
differential equations of special type. In particular, if the
nonlinear medium is endowed still with spatial memory properties,
the propagation of the corresponding waves can be modeled by means
of the so called generalized Ostrovsky evolution equations
\cite{Os}. It is also well known \cite{Wh,MPV,Va} that shortwave
perturbations in a relaxing one dimensional medium can be
described by means of some reduction of the Ostrovsky equations,
coinciding with the Whitham type evolution equation
\begin{equation}
du/dt=2uu_{x}+\int_{\mathbb{R}}\mathcal{K}(x,s)u_{s}ds,  \label{1.1}
\end{equation}%
discussed first in \cite{Wh}. Here the kernel
$\mathcal{K}:\mathbb{R}\times \mathbb{R}\rightarrow \mathbb{R}$
depends on the medium elasticity properties with spatial memory
and can, in general, be a function of the pressure gradient
$u_{x}\in C^{\infty }(\mathbb{R};\mathbb{R}),$ evolving in
respect to  equation (\ref{1.1}). In particular, if $\mathcal{K}(x,s)=%
\frac{1}{2}\mid x-s\mid ,~x,s\in \mathbb{R},$ then equation \ (\ref{1.1}%
) can be reduced to
\begin{equation}
du/dt=2uu_{x}+\partial ^{-1}u,  \label{1.2}
\end{equation}%
which was, in particular, studied before in \cite{Pa,Pa1,Va,MPV,Wh}.

Since some media possess elasticity properties depending strongly on the
spatial pressure gradient $u_{x},~x\in \mathbb{R},$ the corresponding
Whitham kernel looks like
\begin{equation}
\mathcal{K}(x,s):=-\theta (x-s)u_{s}  \label{1.3}
\end{equation}%
for $x,s\in \mathbb{R},$ naturally modeling the relaxing spatial
memory effects. The resulting equation \ (\ref{1.1}) with the
kernel (\ref{1.3}) becomes
\begin{equation}
du/dt=2uu_{x}-\partial ^{-1}u_{x}^{2}:=K[u],  \label{1.4}
\end{equation}%
which appears to possess very interesting mathematical properties.
The latter will be the main topic of the next sections  below.

\section{Lagrangian analysis}

A more mathematically correct form of  equation (\ref{1.4}) looks
like
\begin{equation}
u_{xt}=2(uu_{x})_{x}-u_{x}^{2},  \label{2.1}
\end{equation}%
being a nonlinear hyperbolic flow on the axis $\mathbb{R}$.
Concerning the preceding form of (\ref{1.4}) it is necessary to
define the operation $\partial
^{-1}:C^{\infty }(\mathbb{R};\mathbb{R})\rightarrow C^{\infty }(\mathbb{R};%
\mathbb{R}),$ which is not an easy problem. Since equation
(\ref{2.1}) \ is well defined in the space of ${2\pi }$- periodic
functions $C_{2\pi }^{\infty
}(\mathbb{R};\mathbb{R})$, one can \ determine upon its subspace $\bar{C}%
_{2\pi }^{\infty }(\mathbb{R};\mathbb{R})\subset $ $C_{2\pi }^{\infty }(%
\mathbb{R};\mathbb{R})$ of functions under  the condition $%
\int_{0}^{2\pi }f(s)ds$ $=0$ for any $f\in \bar{C}_{2\pi }^{\infty }(\mathbb{%
R};\mathbb{R})$ the inverse operation
\begin{equation}
\partial ^{-1}(\cdot ):=\frac{1}{2}[\int_{0}^{x}(\cdot )ds-\int_{x}^{2\pi
}(\cdot )ds],  \label{2.2}
\end{equation}%
being definite for all $x\in \mathbb{R}$ and satisfying the natural property
$\partial \cdot \partial ^{-1}=1.$ Thereby, for convenience, we will
consider the flow (\ref{1.4}) as that in the smooth functional submanifold $%
M:=\bar{C}_{2\pi }^{\infty }(\mathbb{R};\mathbb{R}).$ The
corresponding vector field $K:M\rightarrow T(M)$ defines on $M$ a
dynamical system, which appears to possess both Lagrangian and
Hamiltonian properties.

To demonstrate them in detail, consider the partial differential equation (%
\ref{2.1}) and prove that it is of Lagrangian form, that is
\begin{equation}
u_{xt}=-\frac{\delta H_{\vartheta }}{\delta u}:=\xi \lbrack u],  \label{2.3}
\end{equation}%
where $H_{\vartheta }:M\rightarrow \mathbb{R}$ is some Fr\'{e}chet smooth
Lagrangian function. To prove (\ref{2.3}), following the scheme in \cite%
{PM,Ol,FF}, it is enough to state that only    the Volterrian
identity $\xi ^{^{\prime }}=\xi ^{^{\prime }\ast }$ holds, that is
\begin{equation}
\lbrack 2(uu_{x})_{x}-u_{x}^{2}]^{\prime }=[2(uu_{x})_{x}-u_{x}^{2}]^{\prime
\ast },  \label{2.4}
\end{equation}%
where the sign "$^{\prime }$" means the Fr\'{e}chet derivative with respect
to the variable $u\in M$ and $"\ast "$ means the corresponding conjugation
with respect to the natural scalar product on the tangent space $T(M)\simeq
T^{\ast }(M)$. As a result, there exists a Lagrangian function $H_{\vartheta
}:M\rightarrow \mathbb{R}$ in the following explicit form:
\begin{equation}
H_{\vartheta }:=\int_{0}^{2\pi }\mathcal{H}_{\vartheta }dx=\int_{0}^{2\pi
}uu_{x}^{2}dx,  \label{2.5}
\end{equation}%
where we used the standard \cite{PM,Ol} homotopy formula
$H_{\vartheta }=\int_{0}^{1}d\lambda (gradH_{\vartheta }[u\lambda
\},u).$ Thus, expression (\ref{2.3}) can be presented as the Euler
equation
\begin{equation}
\delta \mathcal{L}/\delta u=0,  \label{2.6}
\end{equation}%
where, by definition,
\begin{equation}
\mathcal{L}:=\int_{0}^{t}\int_{0}^{2\pi }(\frac{1}{2}u_{x}u_{\tau }-\mathcal{%
H}_{\vartheta })dxd\tau .  \label{2.7}
\end{equation}%
Recall now, that owing to the standard results \cite{PM,FF,Ol,AM,Ar}, any
Lagrangian system in the form (\ref{2.6}) is Hamiltonian. To show this,
rewrite the action functional (\ref{2.7}) as
\begin{equation}
\mathcal{L}=\int_{0}^{t}[(\varphi ,u_{\tau })-H_{\vartheta }]d\tau ,
\label{2.8}
\end{equation}%
where $\varphi :=(1/2)u_{x}\in T^{\ast }(M)$. Then the condition (\ref{2.6})
gives rise to the equality
\begin{equation}
u_{t}=-\vartheta ~grad~H_{\vartheta }[u]=K[u],  \label{2.9}
\end{equation}%
where, by definition,
\begin{equation}
\vartheta ^{-1}:=\varphi ^{\prime }-\varphi ^{\prime \ast }=\partial
/\partial x.  \label{2.10}
\end{equation}%
As it is easy to see, the operator $\vartheta :=\partial ^{-1}:T^{\ast
}(M)\rightarrow T(M)$ is necessary implectic \cite{PM, FT,FF} and with
respect to the flow (\ref{2.9}) also Noetherian. Thus we have stated the
following \ \cite{PP} theorem.

\begin{theorem}
The partial differential equation (\ref{2.1}) is equivalent on the
functional manifold $M$ \ to the Hamiltonian flow (\ref{2.9}) with the
Hamiltonian function (\ref{2.5}) and co-implectic structure (\ref{2.10}).
\end{theorem}

This result means that our flow (\ref{1.4}) on $M$, being Hamiltonian, is
conservative, thereby one can expect it possesses also an additional hidden
infinite hierarchy of conservation laws, which is very important \cite%
{Ol,PM,FT,No,AS} for its integrability analysis. This assumption, as we
shall show below, appears to hold really.

\section{Gradient-holonomic analysis}

Since any conservation law $\gamma \in D(M)$ satisfies the linear Lax
equation
\begin{equation}
d\psi /dt+K^{\prime \ast }\psi =0,  \label{3.1}
\end{equation}%
where $\psi =grad~\gamma \in T^{\ast }(M)$, under the condition of its
existence in the form of a local functional on $M,$ it can be found for
instance, by means of the asymptotic small parameter method \cite{PM}. In
particular, one easily gets that expressions
\begin{equation}
\psi _{\vartheta }=u_{xx},\quad \psi _{{\eta }_{-1}}=\frac{1}{2}%
(u_{x}^{2}-(u)_{xx}^{2})  \label{3.2}
\end{equation}%
satisfy the Lax equation (\ref{3.1}) and are the gradients of the
corresponding functionals on $M$, that is
\begin{equation}
\psi _{\vartheta }=grad~\gamma _{\vartheta }\quad \psi _{{\eta }%
_{-1}}=grad~\gamma _{{\eta }_{-1}},  \label{3.3}
\end{equation}%
where
\begin{equation}
\gamma _{\vartheta }=\frac{1}{2}\int_{0}^{2\pi }u_{x}^{2}dx\quad \gamma _{{%
\eta }_{-1}}=\frac{1}{2}\int_{0}^{2\pi }uu_{x}^{2}dx.  \label{3.4}
\end{equation}%
Thus, we have stated that our dynamical system (\ref{1.4}) allows additional
invariants (conservation laws), which can be used within the
gradient-holonomic algorithm \cite{PM,MBPS,No} for finding new associated
nontrivial implectic structures on the manifold $M.$ Namely, let us
represent conservation laws (\ref{3.2}) in the scalar product form on $M$ as
\begin{equation}
\gamma _{\vartheta }=(\varphi _{\vartheta },u_{x})\quad \gamma _{{\eta }%
_{-1}}=(\varphi _{{\eta }_{-1}},u_{x}),  \label{3.5}
\end{equation}%
where
\begin{equation}
\varphi _{\vartheta }=\frac{1}{2}u_{x},\quad \varphi _{{\eta }_{-1}}=-\frac{1%
}{2}\partial ^{-1}u_{x}^{2}\in T^{\ast }(M).  \label{3.6}
\end{equation}%
Then operators
\begin{equation*}
\vartheta ^{-1}=\varphi _{\vartheta }^{^{\prime }}-\varphi _{\eta
}^{^{\prime }\ast }=\frac{1}{2}\partial -(-\frac{1}{2}\partial )=\partial ,
\end{equation*}%
\begin{equation}
\eta _{-1}^{-1}=\varphi _{{\eta }_{-1}}^{^{\prime }}-\varphi _{{\eta }%
_{-1}}^{^{\prime }\ast }=\partial ^{-1}u_{xx}+u_{xx}\partial ^{-1}
\label{3.7}
\end{equation}%
will be co-implectic \cite{PM, FF,Ol} on $M,$ and, as it is easy to check,
also Noetherian with respect to our dynamical system (\ref{1.4}). Moreover,
via direct calculations one can show that the corresponding implectic
operators $\vartheta ,~\eta _{-1}:T^{\ast }(M)\rightarrow T(M)$ are
compatible on $M,$ that is for any $\lambda \in \mathbb{R}$ the expression $%
\vartheta +\lambda \eta _{-1}$ is implectic too on $M$ \cite{PM, Ma,FF}. \
Really, it is enough to show \cite{FF,PM} that the operator $\vartheta
^{-1}\eta _{-1}\vartheta ^{-1}:$ $T^{\ast }(M)\rightarrow $ $T(M)$ is
symplectic on $M,$ that is the differential two-form $\Omega
^{(2)}:=\int_{0}^{2\pi }dx(du\wedge \vartheta ^{-1}\eta _{-1}\vartheta
^{-1}du)\in \Lambda ^{2}(M)$ is closed, or $d\Omega ^{(2)}=0.$ The latter
equality is easily checked by direct calculations. This means, in
particular, that all operators of the form
\begin{equation}
\eta _{n}=\vartheta (\eta _{-1}^{-1}\vartheta )^{n}  \label{3.8}
\end{equation}%
for $n\in \mathbb{Z}$ will also be implectic  on $M.$ Another
consequence from this fact is the existence of an infinite
hierarchy of invariants $\gamma _{n}\in D(M),~~n\in \mathbb{Z}$,
satisfying the expressions
\begin{equation}
K[u]=-\eta _{n}~grad~\gamma _{n}.  \label{3.9}
\end{equation}%
As a particular case one can define an implectic operator $\eta :T^{\ast
}(M)\rightarrow T(M)$ in the form
\begin{equation}
\eta =\vartheta \eta _{-1}^{-1}\vartheta =\partial ^{-2}u_{xx}\partial
^{-1}+\partial ^{-1}u_{xx}\partial ^{-2}.  \label{3.10}
\end{equation}%
Whence  from (2.9) we obtain that
\begin{equation}
u_{t}=K[u]=-\vartheta grad~H_{\vartheta }=-\eta grad~H_{\eta },  \label{3.11}
\end{equation}%
where \
\begin{equation*}
H_{\vartheta }=\int_{0}^{2\pi }uu_{x}^{2}dx,\quad H_{\eta }=\int_{0}^{2\pi
}u_{x}^{2}dx.
\end{equation*}%
The set of expressions (\ref{3.8}) can be equivalently rewritten in another
useful form as
\begin{equation}
\lambda \vartheta ~grad~\gamma (\lambda )=\eta ~grad~\gamma (\lambda ),
\label{3.12}
\end{equation}%
being in some sense equivalent \cite{PM,Ol,FF} together with  equation (%
\ref{3.1}) to the adjoint Lax type representation
\begin{equation}
d\Lambda /dt=[\Lambda ,K^{\prime \ast }]  \label{3.12a}
\end{equation}%
for the dynamical system (\ref{1.4}), where  $\Lambda :=\vartheta
^{-1}\eta :T^{\ast }(M)\rightarrow T^{\ast }(M)$ is a so called \cite%
{Ol,FF,PM,FF} recursion operator and $\gamma (\lambda )\in D(M),~\lambda \in
\mathbb{C},$ is a generating function of the infinite hierarchy of
conservation laws (\ref{1.4}). In particular, as $\ |\lambda |\rightarrow
\infty $ the asymptotic expansion
\begin{equation}
grad~\gamma (\lambda )_{{\mid }_{{\mid \lambda \mid }\rightarrow \infty
}}\simeq \sum_{{j}\in \mathbb{Z_{+}}}\lambda ^{-j}~grad~\gamma _{j}
\label{3.13}
\end{equation}%
holds, where
\begin{equation}
grad~\gamma _{n}=\Lambda ^{n}~grad~\gamma _{0},\quad \gamma _{0}:=H_{\eta },
\label{3.14}
\end{equation}%
for all $n\in \mathbb{Z}_{+}.$\ Concerning this infinite hierarchy
of conservation laws one can easily check, that all of them are
dispersiveless. The result obtained above can be  formulated as
the next theorem.

\begin{theorem}
The dynamical system (\ref{1.4}) on the functional manifold $M$ is a
compatible bi-Hamiltonian flow, possessing an infinite hierarchy of
commuting functionally independent dispersionless conservation laws,
satisfying the fundamental gradient identity (\ref{3.12}). The latter is
equivalent together with the relationship (\ref{3.1}) to the adjoint Lax
type representation \ (\ref{3.12a}).
\end{theorem}

As was mentioned above, the hierarchy of commuting flows $K_{n}:=-\vartheta
~grad~\gamma _{n},$ $n\in \mathbb{Z}_{+},$ shows an interesting property of
their dispersionless. In particular, this entails that they can not be
treated effectively by means of the gradient-holonomic algorithm \cite%
{PM,MBPS,No}. In particular, the corresponding asymptotic
solutions to the Lax equations
\begin{equation}
d\varphi /d\tau _{n}+K_{n}^{\prime \ast }\varphi =0,\quad \varphi ^{\prime
}\neq \varphi ^{\prime ^{+}},  \label{3.15}
\end{equation}%
where  $\mid \lambda \mid \rightarrow \infty $  and   $du/d\tau
_{n}=K_{n}[u],~\tau _{n}\in \mathbb{R},$ $n\in \mathbb{Z},$ do not
give rise to explicit functional expressions, defining a new
associated hierarchy of conservation laws for the dynamical system \ (\ref%
{1.4}). Nonetheless, the corresponding hierarchy of dispersive commuting
flows on $M$ does exist for \ (\ref{1.4}), being simply associated with the
trivial flow $du/dt_{0}:=0$ on $M.$ Namely, let $H_{0}\in D(M)$ be a
conservation law of \ (\ref{1.4}), satisfying the kernel condition for the
operator $\eta :T^{\ast }(M)\rightarrow T(M),$ that is
\begin{equation}
du/dt_{0}=0:=\eta ~grad~H_{0}.  \label{3.16}
\end{equation}%
It is easy to find from \ (\ref{3.16}) and \ (\ref{3.10}) that $%
grad~H_{0}=[2(u_{xx})^{-1/2}]_{xx}\in T^{\ast }(M),$ whence
\begin{equation}
H_{0}=4\int_{0}^{2\pi }\sqrt{u_{xx}}dx.  \label{3.17}
\end{equation}%
The obtained invariant \ (\ref{3.17})  allows to construct a new
associated with  (1.4)  commuting flow
\begin{equation}
du/d\tau =-\vartheta ~grad~H_{0}=u_{xxx}(u_{xx})^{-3/2}:=\tilde{K}[u],
\label{3.18}
\end{equation}%
$\tau \in \mathbb{R},$ which, as it is easy to see, already
possesses a nontrivial dispersion. This means that the Lax
equation
\begin{equation}
d\varphi /d\tau +\tilde{K}^{^{\prime }\ast }\varphi =0,  \label{3.19}
\end{equation}%
allows as $|\lambda |\rightarrow \infty $ an asymptotic solution $\ \varphi
:=\varphi (\tau ,x;\lambda )\in T^{\ast }(M)\otimes \mathbb{C},$ where
\begin{eqnarray}
\varphi (\tau ,x;\lambda ) &\simeq &exp(\lambda ^{3}\tau
+\int_{x_{0}}^{x}\sigma (y;\lambda )dy),  \label{3.20} \\
\quad \sigma (x;\lambda ) &\simeq &\sum_{j\in \mathbb{Z_{+}}}\sigma
_{j-1}[u]\lambda ^{-j+1}.  \notag
\end{eqnarray}%
The nontrivial functionals $\gamma _{j-1}:=\int_{0}^{2\pi }\sigma
_{j-1}[u]dx,$ \quad $j\in \mathbb{Z_{+}},$ are, obviously, functionally
independent and commuting conservation laws both of the dynamical system \ (%
\ref{3.18}) and of our dynamical system \ (\ref{1.4}). As a result
of some simple but slightly tedious calculations one finds that
\begin{equation}
\sigma _{-1}=\sqrt{u_{xx}},\text{ \ }\sigma _{0}=\frac{1}{2}%
u_{xx}^{-1}u_{xxx},\text{ \ }\sigma _{1}=\frac{1}{8}%
(u_{xx})^{-5/2}u_{xxx}^{2},...,  \label{3.21}
\end{equation}%
and the\ corresponding hierarchy of already dispersive invariants is given
as\
\begin{equation}
{\gamma _{-1}=\int_{0}^{2\pi }\sqrt{u_{xx}}\;dx,\quad \gamma _{0}=0,\quad }
\label{3.22}
\end{equation}%
\begin{equation*}
{\gamma _{1}=\tfrac{1}{8}\int_{0}^{2\pi }u_{xx}^{-5/2}u_{xxx}^{2}\;dx,}\text{
\ \ }\gamma _{2}=0,\text{{\ }}{\ldots ,}
\end{equation*}%
and so on. Then, owing to conditions \ (\ref{3.18}) and \ (\ref{3.19}) the
generating functional $\gamma (\lambda ):=\int_{0}^{2\pi }\sigma (x;\lambda
)dx,$ \ $\lambda \in \mathbb{C},$ satisfies \cite{PM,FF,MBPS} the following
gradient relationship
\begin{equation}
\lambda ^{2}\vartheta ~grad~\gamma (\lambda )=\eta ~grad~\gamma (\lambda ),
\label{3.23}
\end{equation}%
suitably modifying the relationship \ (\ref{3.12}).

The obtained results are very important for further analytical studying Lax
type integrability of the dynamical system \ (\ref{1.4}) and finding, in
particular, a wide class of its special soliton like and quasi-periodic
solutions by means of analytical quadratures. Some of these aspects of the
integrability problem are presented in the section below.

\section{Lax type representation and finite dimensional reductions}

Since the functional solution \ (\ref{3.20}) satisfies the Lax type equation
\ (\ref{3.19}), it can be considered \cite{MBPS,No,FT} as a Bloch type
eigenfunction of the adjoint Lax type representation \ \ (\ref{3.12a}), that
is
\begin{equation}
\Lambda \varphi (x;\lambda )=\lambda ^{2}\varphi (x;\lambda )  \label{4.1}
\end{equation}%
for all $\lambda \in \mathbb{C}$ and $x\in \mathbb{R}.$\bigskip\ This gives
rise, following the gradient-holonomic algorithm \cite{MBPS,No}, to the
existence of a standard Lax type representation for the associated dynamical
system \ (\ref{3.18}) and, thereby, for our Whitham type dynamical system \ (%
\ref{1.4}). Omitting here the related calculations, we find surprisingly
that this adjoint to \ (\ref{4.1}) Lax type spectral problem for the flow \ (%
\ref{1.4}) is equal to
\begin{equation}
Lf:=\left(
\begin{array}{cc}
-i\lambda & \lambda (u_{xx}-1) \\
-\lambda & i\lambda%
\end{array}%
\right) f,  \label{4.2}
\end{equation}%
where an eigenfunction $f\in L_{\infty }(\mathbb{R};\mathbb{C}^{2})$ and $%
\lambda \in \mathbb{C}$ is a time independent spectral parameter. The result
\ (\ref{4.2}) can be effectively used for solving our nonlinear equation \ (%
\ref{1.4}), making use either of the inverse spectral transform method \cite%
{No,DMN,MBPS,AS,FT} or of the dual Bogoyavlensky-Novikov method \cite{No,PM}
of finite dimensional reductions. For the latter case we need to construct
an  finite dimensional  invariant symplectic functional submanifolds $%
M^{2N}\subset M,$ $N\in \mathbb{Z}_{+},$ and to represent \ the
main vector fields $d/dx$ \ and \ $d/dt$ on them as the
corresponding \bigskip commuting to each other Hamiltonian flows.
Moreover, \ since these flows on $\ M^{2N}$ appear to be
Liouville-Arnold integrable, we obtain both the complete
integrability of our dynamical system \ (\ref{1.4}) in quadratures
and their exact solutions, expressed, in general, by means of
Riemannian theta-functions \cite{No,DMN,MBPS,FT} on some specially
constructed algebraic Riemannian surfaces.

Below we consider, for simplicity,  the following invariant two-
and
four-dimensional functional submanifolds:%
\begin{equation}
i)\text{ \ \ \ }M^{2}:=\{u\in M:grad\mathcal{L}_{2}[u]=0\},  \label{4.3}
\end{equation}%
where $\mathcal{L}_{2}:=H_{\vartheta }+c_{\eta }H_{\eta }\in \mathcal{D}(M),$
and
\begin{equation}
ii)\text{ \ \ \ }M^{4}:=\{u\in M:grad\mathcal{L}_{4}[u]=0\},  \label{4.4}
\end{equation}%
where $\mathcal{L}_{4}:=\gamma _{-1}+c_{\vartheta }H_{\vartheta }+c_{\eta
}H_{\eta }\in \mathcal{D}(M).$

\textbf{Case i). }We have, therefore \cite{PM,Di,No,DMN}, on the invariant
manifold $M^{2}$ commuting Hamiltonian vector fields $\ \ d/dx$ and $d/dt$ \
with respect to\bigskip\ the canonical symplectic structure
\begin{equation}
\omega ^{(2)}:=d\alpha ^{(1)},  \label{4.5}
\end{equation}%
where 1-form $\alpha ^{(1)}\in \Lambda ^{1}(M)$ is determined by the
Gelfand-Dickey \cite{Di,PM} relationship%
\begin{equation}
d\mathcal{L}_{2}[u]=grad\mathcal{L}_{2}[u]du+d\alpha ^{(1)}/dx,  \label{4.6}
\end{equation}%
holding on $M$. One now easily finds, that for all $u\in M^{2}$ $\subset M$%
\begin{eqnarray}
grad\mathcal{L}_{2}[u] &=&u_{x}^{2}-2(u_{x}u)_{x}-2c_{\eta }u_{xx}=0,
\label{4.7} \\
\alpha ^{(1)} &=&2(u+c_{\eta })u_{x}du,  \notag \\
\omega ^{(2)} &=&d[2(u+c_{\eta })u_{x}]\wedge du:=dp\wedge dq,  \notag
\end{eqnarray}%
where we have put $p:=2(u+c_{\eta })u_{x}$ and $q:=u.$ \ The corresponding
Hamiltonian functions $h^{(x)}$ and $h^{(t)}\in \mathcal{D}(M^{2})$ for
Hamiltonian flows
\begin{eqnarray}
dq/dx &=&\partial h^{(x)}/\partial p,\text{ \ \ \ \ }dp/dx=-\partial
h^{(x)}/\partial q,  \label{4.8} \\
dq/dt &=&\partial h^{(t)}/\partial p,\text{ \ \ \ \ }dp/dt=-\partial
h^{(t)}/\partial q  \notag
\end{eqnarray}%
are found \cite{PM}, respectively, from the determining relationships
\begin{equation}
grad\mathcal{L}_{2}[u]\text{ }u_{x}:=-dh^{(x)}/dx,\text{ \ \ \ \ }grad%
\mathcal{L}_{2}[u]\text{ }u_{t}:=-dh^{(t)}/dx,  \label{4.9}
\end{equation}%
whence we get that
\begin{equation}
h^{(x)}=(u+c_{\eta })u_{x}=\frac{p^{2}}{4(q+c_{\eta })},\text{ \ \ \ \ }%
h^{(t)}=-2c_{\eta }(u+c_{\eta })u_{x}=\frac{-c_{\eta }p^{2}}{2(q+c_{\eta })}%
\text{\ .}  \label{4.10}
\end{equation}%
One sees easily that two flows $d/dx$ and $d/dt$ on the
two-dimensional invariant submanifold $M^{2}\subset M$ \ \ of \
infinite period are proportional, confirming the classical fact\
\cite{Ar,AM}: upon two-dimensional symplectic manifold there exist
the only functionally-independent invariant commuting to each
other.\ \ \ \

The set of Hamiltonian equations \ (\ref{4.9}) \ \ \ \ for the flows $d/dx$
and $d/dt$ has the simple form
\begin{eqnarray}
\frac{dq}{dx} &=&\frac{p}{2(q+c_{\eta })},\text{ \ \ \ }\frac{dp}{dx}=\frac{%
p^{2}}{4(q+c_{\eta })^{2}},  \label{4.11} \\
\frac{dq}{dt} &=&\frac{-c_{\eta }p}{(q+c_{\eta })},\text{ \ \ \ }\frac{dp}{dt%
}=\frac{-c_{\eta }p^{2}}{2(q+c_{\eta })^{2}},  \notag
\end{eqnarray}%
whose solution, given by the exact formula
\begin{equation}
q(x,t)=-c+[\frac{3}{2}\sqrt{\bar{h}^{(x)}}(x-2c_{\eta }t)+\bar{k}%
]^{2/3}\Rightarrow u(x,t),  \label{4.12}
\end{equation}
with $\bar{k}\in \mathbb{R}$ being some real constant, supplies us,
evidently, with an exact partial one-parametric solution to our Whitham type
nonlinear equation \ (\ref{1.4}).

\bigskip \textbf{Case ii). }Similarly as above, we find the quantities
\begin{eqnarray}
grad\mathcal{L}_{4}[u] &=&(\frac{1}{2\sqrt{u_{xx}}})_{xx}+c_{\vartheta
}[u_{x}^{2}-2(uu_{x})_{x}]-2c_{\eta }u_{xx}=0,  \label{4.13} \\
\alpha ^{(1)} &=&[2(c_{\vartheta }u+c_{\eta })u_{x}-(\frac{1}{2\sqrt{u_{xx}}}%
)_{x}]du+\frac{1}{2\sqrt{u_{xx}}}du_{x},  \notag
\end{eqnarray}%
whence the symplectic structure is given as
\begin{eqnarray}
\omega ^{(2)} &=&d[2(c_{\vartheta }u+c_{\eta })u_{x}-(\frac{1}{2\sqrt{u_{xx}}%
})_{x}]\wedge du+d(\frac{1}{2\sqrt{u_{xx}}})\wedge du_{x}  \label{4.14} \\
&=&dp_{1}\wedge dq_{1}+dp_{2}\wedge dq_{2},  \notag
\end{eqnarray}%
where we put, by definition, $(q_{1}:=u,$ $q_{2}:u_{x},$ $\
p_{1}:=2(c_{\vartheta }u+c_{\eta })u_{x}-(\frac{1}{2\sqrt{u_{xx}}})_{x},$ \ $%
p_{2}:=\frac{1}{2\sqrt{u_{xx}}})\in M^{4}$, which are canonical
symplectic coordinates on the invariant functional submanifold
$M^{4}\subset M.$ The commuting to each other Hamiltonian
functions, related with  flows $d/dx$ and \ $d/dt$, are equal to
the next algebraic expressions
\begin{eqnarray}
h^{(x)} &=&u_{x}^{2}(c_{\vartheta }u+c_{\eta })-(\frac{1}{2\sqrt{u_{xx}}}%
)_{x}u_{x}-\sqrt{u_{xx}}  \label{4.15} \\
&=&3q_{2}^{2}(c_{\vartheta }q_{1}+c_{\eta )}-q_{2}p_{1}-1/(2p_{2})  \notag
\end{eqnarray}%
and
\begin{equation}
h^{(t)}=q_{1}/(2p_{2})-2q_{1}q_{2}[p_{1}-q_{2}(c_{\vartheta }q_{1}+c_{\eta
})].  \label{4.16}
\end{equation}%
As a result, we \ have reduced our Whitham type dynamical system \ (\ref{1.4}%
)\ \ upon the constructed four-dimensional invariant submanifold $%
M^{4}\subset M,$ on which it is exactly equivalent to two commuting
canonical Hamiltonian flows
\begin{eqnarray}
dq_{j}/dx &=&\partial h^{(x)}/\partial p_{j},\text{ \ \ \ }%
dp_{j}/dx=-\partial h^{(x)}/\partial q_{j,}  \notag \\
dq_{j}/dt &=&\partial h^{(t)}/\partial p_{j},\text{ \ \ \ }%
dp_{j}/dt=-\partial h^{(t)}/\partial q_{j}  \label{4.17}
\end{eqnarray}%
for $j=\overline{1,2},$ where the corresponding Poisson bracket $%
\{h^{(x)},h^{(t)}\}=0$ on $M^{4}.$ Thereby, owing to the classical
Liouville-Arnold \ theorem \cite{Ar,AM,PM} our Whitham type dynamical system
\ (\ref{1.4}), reduced invariantly upon the four-dimensional invariant
submanifold $M^{4}\subset M,$ is completely integrable by quadratures. This
result we will formulate as a final theorem.

\begin{theorem}
The Whitham type dynamical system \ (\ref{1.4}), reduced upon the
invariant two-parametric four-dimensional functional submanifold \
$M^{4}\subset M$ is exactly equivalent to the set of two commuting
to each other canonical Hamiltonian flows \ (\ref{4.17}), which
are completely Liouville-Arnold integrable by quadratures systems.
The corresponding  Hamiltonian functions are given by expressions
\ (\ref{4.15}) and \ (\ref{4.16}).
\end{theorem}

The results obtained above make it possible to construct a wide class of
exact two-parametric solutions of the Whitham type nonlinear equation \ (\ref%
{1.4}) by means of quadratures. This very interesting and
important problem we plan to investigate in detail in another
paper.

\section{Regularization scheme and the integrability problem}

Define a smooth periodic unction $v\in C_{2\pi }^{\infty }(\mathbb{R};%
\mathbb{R}),$ such that%
\begin{equation}
v:=\partial ^{-1}u_{x}^{2}  \label{5.1}
\end{equation}%
for any $x,t\in \mathbb{R},$ where the function \bigskip $u\in C_{2\pi
}^{\infty \text{ }}(\mathbb{R};\mathbb{R})$ solves  equation \ (\ref{1.1}%
). Then it is easy to state that the following nonlinear dynamical system
\begin{equation}
\left.
\begin{array}{c}
u_{t}=2uu_{x}-v \\
v_{t}=2uv_{x}%
\end{array}%
\right\} :=K[u,v]  \label{5.2}
\end{equation}%
of hydrodynamic type, being already well defined on the extended $2\pi $%
-periodic functional space $\mathcal{M}:=C_{2\pi }^{\infty }(\mathbb{R};%
\mathbb{R}),$ is completely equivalent to that given by expression (\ref{1.1}%
). Thereby,  mapping \ (\ref{5.1}) regularizes the previously \
not completely determined expression (\ref{1.1}), making it
possible to pose a new integrability problem for dispersionless dynamical system \ (\ref%
{5.2}) of hydrodynamic type in the functional space $\mathcal{M}.$
To proceed, one can analyze this problem by means of the
gradient-holonomic method \cite{PM, MBPS}, which is similar to the
approach applied above for studying dynamical system \
(\ref{1.1}). Based on preliminary performed calculations and
obtained analytical properties of \ dynamical systems \
(\ref{5.2}) we can state that it is also an integrable flow in the
functional space $\mathcal{M},$ possessing a suitable Lax type
representation. We plan to investigate this problem in detail in
another work under preparation.

\section*{\protect\bigskip Acknowledgments}

Two of the authors (A.P. and N.B.) are cordially indebted to the
Abdus Salam International Centre for Theoretical Physics, Trieste,
Italy, for the hospitality during their ICTP-2007 research
scholarships. Their warm thanks are directed in particular to
Professor L\'{e} Dung Tr\'{a}ng, Head of the Mathematical
Department, for an invitation to visit ICTP, where the creative
atmosphere was instrumental in the completion of this work. \ One
of the authors (A.P.) expresses his appreciation  to Prof. B.A.
Dubrovin (SISSA, Trieste, Italy) for kind hospitality at SISSA,
Trieste, the ESF-2006 Research Program, within which some results
of the article were obtained.

\end{document}